\documentclass{elsarticle}
\usepackage[usenames,dvipsnames]{color}
\usepackage{algorithm}
\usepackage{algorithmic}
\usepackage{graphicx}
\usepackage{amsmath}

\begin{document}
\begin{frontmatter}
\title{
GPU accelerated Monte Carlo simulation \\of Brownian motors dynamics with CUDA} 
\author{J. Spiechowicz}
\author{M. Kostur}
\author{L. Machura\corref{cor1}}
\ead{lukasz.machura@us.edu.pl}
\address{Institute of Physics, University of Silesia, 40-007 Katowice, Poland}
\cortext[cor1]{corresponding author: tel. +48 32 359 2107; fax. +48 32 258 3653}
\begin{abstract}
This work presents an updated and extended guide on methods of a proper acceleration of the Monte Carlo integration of stochastic differential equations with the commonly available NVIDIA Graphics Processing Units using the CUDA programming environment. We outline the general aspects of the scientific computing on graphics cards and demonstrate them with two models of a well known phenomenon of the noise induced transport of Brownian motors in periodic structures. As a source of fluctuations in the considered systems we selected the three most commonly occurring noises: the Gaussian white noise, the white Poissonian noise and the dichotomous process also known as a random telegraph signal. The detailed discussion on various aspects of the applied numerical schemes is also presented. The measured speedup can be of the astonishing order of about 3000 when compared to a typical CPU. This number significantly expands the range of problems solvable by use of stochastic simulations, allowing even an interactive research in some cases. 
\end{abstract}
\begin{keyword}
Stochastic differential equation, Langevin equation, graphics processing unit, GPGPU, NVIDIA, CUDA, numerical simulation, Monte Carlo method, Brownian motor, Gaussian noise, Poissonian noise, dichotomous noise
\end{keyword}
\end{frontmatter}
{\bf PROGRAM SUMMARY} \\
\begin{small}
\noindent
{\em Manuscript Title:} GPU accelerated Monte Carlo simulation of Brownian motors \linebreak dynamics with CUDA \\
{\em Authors:} Jakub Spiechowicz, Marcin Kostur and {\L}ukasz Machura \\
{\em Program Title:} \emph{poisson}, \emph{dich} \\
{\em Journal Reference:} \\
{\em Catalogue identifier:} \\
{\em Licensing provisions:} LGPLv3 \\
{\em Programming language:} CUDA C \\
{\em Computer:} Any with CUDA-compliant GPU \\
{\em Operating system:} No limits (tested on Linux) \\
{\em RAM:} Hundreds of megabytes for typical case \\
{\em Number of processors used:} Single graphics processing unit \\
{\em Keywords:} Stochastic differential equation, Langevin equation, graphics processing unit, GPGPU, NVIDIA, CUDA, numerical simulation, Monte Carlo method, Brownian motor, Gaussian noise, Poissonian noise, dichotomous noise\\
{\em PACS:} 05.10.Gg, 05.40.-a, 05.40.Ca, 05.40.Jc, 05.60.Cd, 05.60.-k \\
{\em Classification:} 4.3, 23 \\
{\em External routines/libraries:} The NVIDIA CUDA Random Number Generation library (cuRAND) \\
{\em Nature of problem:} Graphics processing unit accelerated numerical simulation of stochastic differential equation \\
{\em Solution method:} The jump-adapted simplified weak order 2.0 predictor-corrector method is employed to integrate the Langevin equation of motion. Ensemble-averaged quantities of interest are obtained through averaging over multiple independent realizations of the system generated by means of Monte Carlo method \\
{\em Unusual features:} The actual numerical simulation run exclusively on graphics processing unit using the CUDA environment. This allows for a speedup as large as about 3000 when compared to a typical CPU. \\
{\em Running time:} a few seconds\\
\end{small}
%
%
\hspace{1pc}
\section{Introduction}
\label{sec1}
The stochastic dynamics has been a very useful tool for analysis of problems in various areas of science, engineering and finance \cite{gardiner}. It takes into consideration the indispensable fluctuations of the system which often play a dominant role in explanation of the occurring phenomena. A remarkable example of this situation is Brownian particle dynamics \cite{reimann2002,hanggi2009,julicher1997,kay2007}.

One of the approaches that have been developed for analysis of dynamical systems subjected to noise is formulated in terms of a stochastic differential equation (SDE) \cite{hanggi1982} that specifies the evolution of stochastic process. Similarly to the case of ordinary differential equations, they are usually rather intractable analytically and explicit solutions are given only for very few cases \cite{gardiner}. However, from a computational point of view, stochastic differential equations can be implemented on modern personal computers without much effort. Moreover, when the dimensionality of the problem is greater than three it is often the only applicable numerical method. In practice, to obtain statistical quantities of interest one has to run a large number of realizations of stochastic differential equation and take average over all of these paths. Each of them is generated by means of Monte Carlo method \cite{binder}. This proceeding is simpler and often much faster than other approaches that have been developed so far for studying such systems. It is due to the recent evolution of computer architectures towards multiprocessor platforms. 

In particular, the advent of Compute Unified Device Architecture (CUDA) \cite{cuda}, a general purpose parallel computing architecture of modern NVIDIA graphics processing units (GPUs) has taken the computational science into completely new level of possibilities \cite{januszewski2009, seibert2011, barros2011, polyakov2013, januszewski2014}. Today, the latest commodity of NVIDIA GPUs are capable of performing trillions of floating point operations per second as a result of employing the power of massively parallel architecture with hundreds of cores on a single silicon chip. In order to take advantage of such hardware one has to carefully redesign the algorithms. The ideal situation occurs when the problem inherently decomposes into a large number of independent tasks whose results can be combined in a simple way, e.g. ensemble averaging in Monte Carlo simulation of stochastic differential equation \cite{kloeden1992, platen2010}. It is an example of a so called "embarrassingly parallel problem" that may particularly benefit from a parallel workload. It has been a few years since the last article devoted to the subject of numerical solution of stochastic differential equations by harvesting the power dormant in the modern GPU was published \cite{januszewski2009} and therefore we decided to present an updated and extended step by step guide how to properly accelerate the stochastic dynamics simulations.

This paper is organized as follows: first, we briefly introduce the CUDA environment. Next, a short overview of the discussed Brownian dynamics models is presented. Then, the algorithms employed to numerically solve these problems are laid out. This is followed by the guide how to properly accelerate numerical solution of stochastic differential equations with CUDA. Next, programs validations and performance measurements on GPU and CPU hardware are presented, both in single and double precision. The final section contains summary and some conclusions. We also provide the source code of example programs which demonstrate the techniques described in the paper.
\section{A primer on CUDA}
\label{sec2}
The name CUDA often embodies not only the hardware architecture of the GPU but also the software which is used to access it. We will now briefly outline the architecture of CUDA compatible devices as well as the abstraction layer between the programmer and the hardware. 

NVIDIA distinguishes CUDA GPUs by their Compute Capability \cite{cuda}. Devices of a greater Compute Capability offer more advanced features and better performance. For example, GPUs with Compute Capability 1.3 and higher support double precision floating point arithmetic \cite{cuda}. From the hardware viewpoint, CUDA is built on the concept of a streaming multiprocessor (MP). Such a multiprocessor is a SIMT (Single Instruction, Multiple Threads) unit formed of several scalar processors (SPs) which are capable of executing a certain job. MPs usually execute a whole warp of threads at once. The warp size specifies the number of threads within a warp. The time it takes to execute one warp depends on what kind of operation the threads need to perform. Moreover, each MP has four types of limited on-chip memory: a set of 32-bit registers, a shared memory block, a constant and texture cache. Since a scalar processor can only access its own registers they cannot be used to share the data between threads. However, the other types of memory are common to all SPs and therefore can be utilized to do it.
\begin{figure}[h]
    \centering
    \includegraphics[width=1.0\linewidth]{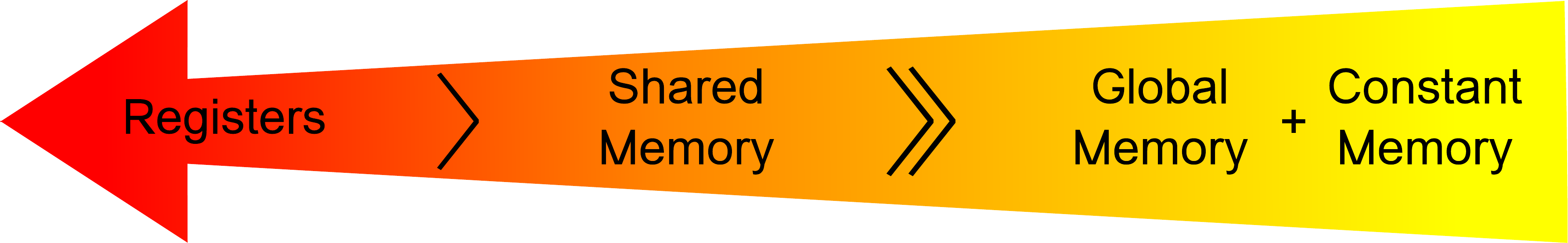}
    \caption{The schematic representation of the GPU memory hierarchy. The above arrow points in the direction of ever-faster types of memory. Its width illustrates the available size of particular type of memory.}
    \label{fig1}
\end{figure}

Probably the most important feature of the CUDA architecture which often determines whether the program is optimized or not is the memory hierarchy \cite{cuda}. The crucial difference between its types is access times. What matters most is how close they are to the actual host memory. A rule of thumb is: the bigger the size of a certain kind of memory, the more time it takes to access it. The slowest type of memory is the global device memory of the GPU. It is the main memory of the graphic card that can be read and written from both the host and device. The name global suggest that it is available to all threads launched on the device and persists for the lifetime of the application. However, global memory accesses are latency expensive operations, taking several hundred clock cycles of the GPU to complete. The fastest kind of memory currently available on graphic cards is the previously mentioned set of registers. Generally, accessing a register consumes zero extra clock cycles per instruction, but delays may occur due to register read-after-write dependencies and register memory bank conflicts. Maximum number of 32-bit registers per thread is presently limited to 255 \cite{cuda}. If there are insufficient registers, data is stored in a local memory. However, here the adjective "local" refers to scope not physical locality. It means that its access time is comparable to the global memory and therefore data that cannot be stored in the set of registers should be loaded into the shared memory which in terms of speed stands in between the global memory and the registers.

The abstraction layer between programmer and the device is built around the concept of a kernel. It is a part of the source code that is executed in parallel by the scalar processors. Kernels are invoked as bunches of threads arranged in one-, two- or three-dimensional blocks. Each block is assigned to a unique multiprocessor. They are further organized into a one- or two-dimensional grid. The composition of the grid and blocks is determined at the time of kernel invocation. Each parallel thread has its unique ID which can be calculated assuming knowledge of the grid and block structure. Threads that reside in a single multiprocessor or equivalently block can exchange data between themselves through the previously mentioned shared memory. Another important feature which is available for the block of threads is an execution synchronization.

The CUDA also provides a comprehensive development environment for C developers building GPU-accelerated applications. It includes a CUDA C compiler, math libraries, and tools for debugging and optimizing the performance of applications. CUDA C is a simple extension of the C programming language which includes several new keywords and expressions that make it possible to distinguish between host and device functions as well as transfer data between them.

\section{Noise induced transport in periodic structures}
We now present two models of Brownian particle dynamics which we concentrate upon later and are of particular interest in many disciplines of science. It is already a well known fact that one can obtain a measurable particle current in periodic structures without application of any external biasing force or field gradient \cite{hanggi1996, astumian2002, hanggi2005}. The consideration of such unusual noise induced transport phenomenon originated from biology and stimulated many scientists working in diverse areas of physics, chemistry and biology itself. As a first approximation of real fluctuations with a short correlation time, Gaussian white noise has been predominantly used. Such additive white noise is not capable of inducing a finite current due to the laws of equilibrium dynamics \cite{risken}. However, there are other stochastic processes frequently met in physical, chemical or biological context which are able to do it due to their inherent statistical asymmetry \cite{luczka1995, hanggi1996re}.

A prominent example of such non-equilibrium fluctuations is a shot noise. Its realizations consist of sequences of very sharp pulses with random heights and randomly distributed times between subsequent pulses. A particular mathematical model of shot noise is formulated as white Poissonian noise \cite{spiechowicz2013, spiechowiczarxiv, spiechowicz2014}. It is a sequence of $\delta$-shaped pulses which occur at times forming Poissonian point process. The amplitudes of the pulses are independent random variables which are distributed according to a common probability density. Such white Poissonian noise commonly occurs in various micro-structures \cite{czernik1997}. This non-equilibrium process is statistically asymmetric in the sense that its cumulants of odd order do not vanish and as a consequence it may induce a directed transport. \cite{luczka1995, hanggi1996re}.

\subsection{Model I}
As the first model let us consider an over-damped Brownian particle which is driven by white Poissonian noise in a spatially periodic potential. The dynamics of the system is described by Langevin equation of the form \cite{czernik1997}
\begin{equation}
    \label{eq1}
    \dot{x}(t) = -V'(x) + \eta(t) + \xi(t),
\end{equation}
where $V(x)$ is a periodic potential with period $L$, so $V(x + L) = V(x)$. It is assumed to be symmetric piece-wise linear potential, namely
\begin{equation}
    \label{eq2}
    V(x) = \left\{ \begin{array}{ll} 1+x, & x \in [-1,0)\, \textrm{mod} \, L\\ 1-x, & x \in [0,1]\, \textrm{mod} \, L.\\ \end{array} \right.
\end{equation}
\begin{figure}[h]
    \centering
    \includegraphics[width=0.45\linewidth]{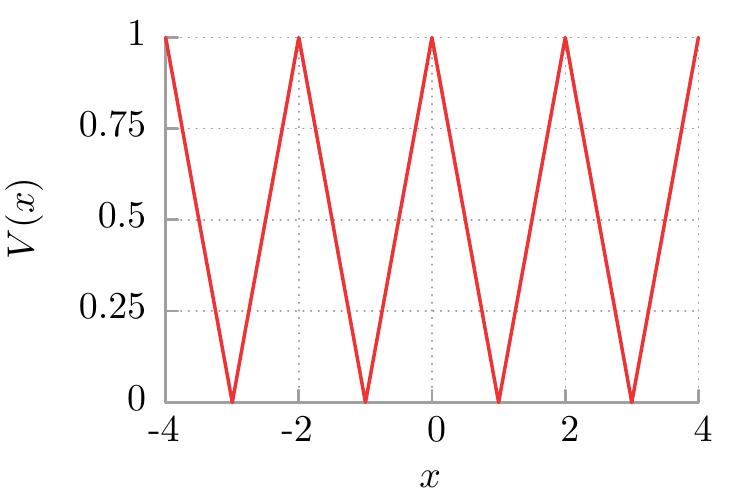}
    \caption{Periodic structure in the form of \emph{symmetric} piece-wise linear potential of period $L = 2$.}
    \label{fig2}
\end{figure}
The process $\eta(t)$ is white Poissonian noise defined as
\begin{equation}
    \label{eq3}
    \eta(t) = \sum_{i=1}^{n(t)}z_i \delta(t - t_i) - a.
\end{equation}
The random times $t_i$ form a Poisson sequence, i.e., the probability that a sequence of $k$ impulses occurs in the interval $(0,t)$ is given by the Poisson distribution
\begin{equation}
    \label{eq4}
    \mbox{Pr}\{n(t) = k\} = \frac{(\lambda t)^k}{k!}e^{-\lambda t}.
\end{equation}
The parameter $\lambda$ determines the average number of $\delta$-pulses per unit time. The amplitudes $\{z_i\}$ are distributed according to exponential probability density
\begin{equation}
    \label{eq5}
    \rho(z) = \zeta^{-1} \theta(z) \exp{(-z/\zeta)},
\end{equation}
where $\theta(z)$ denotes the Heaviside step function. As a consequence the mean value $\langle z \rangle = \zeta$. The quantity
\begin{equation}
    \label{eq6}
    a = \lambda \langle z \rangle
\end{equation}
describes according to (\ref{eq3}) the negative valued bias of the white Poissonian noise realization between consecutive $\delta$-pulses. The process $\eta(t)$ is therefore of zero mean and possesses auto-correlation function  
\begin{equation}
    \label{eq7}
    \langle \eta(t) \rangle = 0, \quad \langle \eta(t)\eta(s) \rangle = 2\lambda\zeta^2\delta(t-s).
\end{equation}
The last equation defines the Poissonian noise intensity 
\begin{equation}
    \label{eq8}
    D_P = \lambda\zeta^2.
\end{equation}
Thermal equilibrium fluctuations due to the coupling of the particle with the environment are modeled as usual by unbiased Gaussian white noise
\begin{equation}
    \label{eq9}
    \langle \xi(t) \rangle = 0, \quad \langle \xi(t)\xi(s) \rangle = 2k_BT\delta(t-s),
\end{equation}
where $k_B$ is the Boltzmann constant and $T$ is the temperature. The noise intensity factor
\begin{equation}
    \label{eq10}
    D_G = k_BT
\end{equation}
follows from the fluctuation-dissipation theorem \cite{luczka1999}. Furthermore, we assume that thermal fluctuations $\xi(t)$ are uncorrelated with non-equilibrium noise $\eta(t)$, so
\begin{equation}
    \label{eq11}
    \langle \xi(t)\eta(s) \rangle = \langle \xi(t) \rangle \langle \eta(s) \rangle = 0.
\end{equation}

\subsection{Model II}
Along with the white Poissonian noise, the dichotomous process, known also as the random telegraph signal \cite{kula1998prl, kula1998, kostur2001}, is an another important type of a non-equilibrium noise that has wide applications in various fields of science. In particular, it may provide a realistic representation of an actual physical situation such as random transitions between meta-stable configurations in a two level system, the phenomenon which is a paradigm for numerous theories. Its realizations consist of randomly distributed jumps between the two states. Being important example of colored noise it mimics the effects of the finite correlation time of the real fluctuations \cite{luczka2005}.

The second model we analyze is that of an over-damped Brownian particle which is subjected to the dichotomous noise in a periodic structure. Consequently, the dynamics of the system is given by the following equation \cite{kula1998prl}
\begin{equation}
    \label{eq12}
    \dot{x}(t) = -V'(x) + \varGamma(t) + \xi(t).
\end{equation}
Here $V(x)$ and $\xi(t)$ are the previously defined the symmetric piece-wise linear potential and the Gaussian white noise, respectively. The process $\varGamma(t)$ represents non-equilibrium fluctuations modeled by the dichotomous noise
\begin{equation}
    \label{eq13}
    \varGamma(t) = \{-F_a,F_b\}, \quad F_a,F_b > 0,
\end{equation}
with the transition probabilities
\begin{equation}
    \label{eq14}
    \mbox{Pr}(-F_a \to F_b) = \mu_a = \frac{1}{\tau_a}, \quad \mbox{Pr}(F_b \to -F_a) = \mu_b = \frac{1}{\tau_b},
\end{equation}
where $\tau_a$ and $\tau_b$ are the mean waiting times in states $F_a$ and $F_b$, respectively. The process $\varGamma(t)$ possesses the auto-correlation
\begin{equation}
    \label{eq15}
    \langle \varGamma(t)\varGamma(s) \rangle = \frac{D_D}{\tau}e^{-|t-s|/\tau},
\end{equation}
with the intensity $D_D$ and the correlation time $\tau$ determined by
\begin{equation}
    \label{eq16}
    D_D = \frac{\mu_a\mu_b(F_a + F_b)^2}{(\mu_a + \mu_b)^3}, \quad \frac{1}{\tau} = \frac{1}{\tau_a} + \frac{1}{\tau_b} = \mu_a + \mu_b.
\end{equation}
We further assume that $F_a\mu_b = F_b\mu_a$. Then
\begin{equation}
    \label{eq17}
    \langle \varGamma(t) \rangle = 0, \quad D_D = F_aF_b\tau.
\end{equation}
If $F_a = F_b$ the dichotomous noise is symmetric. Otherwise it is asymmetric. As usual, we assume that $\varGamma(t)$ is not correlated with $\xi(t)$.

The most prominent transport quantity for the above systems is the average velocity $\langle v \rangle$. In the asymptotic long time limit it can be defined as
\begin{equation}
    \label{eq18}
    \langle v \rangle = \lim_{t \to \infty} \frac{\langle x(t) \rangle}{t}.
\end{equation}
The corresponding stationary probability density $p(y)$ is expressed in the following way
\begin{equation}
    \label{eq19}
    p(y) = \lim_{t \to \infty} \langle \delta\left(y - x(t)\right) \rangle.
\end{equation}
Generally, this steady state carries a probability current $J$. It is significant that both considered systems can actually be studied analytically in the limiting case of no coupling between the particle and the environment, so when $D_G = k_BT = 0$. The stationary probability densities and currents satisfy the following differential equations which was obtained in \cite{luczka1995} and \cite{kula1996} for the first and the second model, respectively:
\begin{subequations}
    \label{eq20}
    \begin{align}
        J_I &= \zeta[-V'(x) - \lambda\zeta]p'(x) + [-V'(x) - \zeta V''(x)]p(x), \\
        J_{II} &= - \frac{D_{eff(x)}}{1 - \tau V''(x)}p'(x) - \frac{V'(x) + D'_{eff}(x)}{1 - \tau V''(x)}p(x),
    \end{align}
\end{subequations}
where the prime denotes a derivative with respect to $x$ and the effective diffusion function $D_{eff}(x) = \tau[F_a + V'(x)][F_b - V'(x)]$. The value of $J$ is obtained by the periodic boundary condition $p(-L/2) = p(L/2)$ and normalization $\int_{-L/2}^{L/2} dx \,p(x) = 1$. We remark that for the piece-wise linear potential the current can be calculated explicitly. For instance, when $L = 2$ it reads
\begin{subequations}
    \label{eq21}
    \begin{align}
        J_I &= \frac{1}{4D_P}\frac{\alpha_+ - \alpha_-}{(\alpha_+ - 1)(\alpha_- - 1)}, \\
        J_{II} &= \frac{1}{4D_D}\frac{\beta_+ - \beta_-}{(\beta_+ - 1)(\beta_- - 1)},
    \end{align}
\end{subequations}
with
\begin{equation}
    \label{eq22}
    \alpha_{\pm} = \exp\left( \frac{a}{D_P(a \mp 1)} \right) \quad \beta_{\pm} = \exp\left( \frac{1}{(F_a \pm 1)(F_b \mp 1)\tau} \right).
\end{equation}
We end this section with the presentation of the relationship between the stationary average velocity $\langle v \rangle$ which can be obtained from estimate of the moment $\langle x(t) \rangle$ by using (\ref{eq18}) and the steady state current $J$
\begin{equation}
    \label{eq23}
    \langle v \rangle = \langle \dot{x} \rangle = \langle -V'(x) \rangle = - \int_{0}^{L} dx \,V'(x)p(x) = LJ
\end{equation}
where the expression for $V'(x)p(x)$ has been inserted from (\ref{eq20}) and periodicity of $p(x)$ and $V(x)$ has been utilized. We stress that this relation is not so simple for the case of non-stationary states.
\section{Numerical solution of stochastic differential equations}
\label{sec3}
This section contains a detailed presentation of algorithms employed to solve the models given by (\ref{eq1}) and (\ref{eq12}). Since stochastic differential equations describing real problems usually cannot be tackled analytically, direct numerical simulations have to be carried to obtain quantities of interest. There is an abundance of methods for solving stochastic differential equations. At the present moment two books written by Platen et al. \cite{kloeden1992, platen2010}. are the most comprehensive references in this field. Most numerical algorithms use a stochastic Taylor expansion technique based on a discrete time approximation. They are classified as either strong or weak schemes. Methods that generate good approximation of a probability distribution and its moments are considered weak schemes. In the remaining part of this paper we present only such class of algorithms.

In this section, we will consider for simplicity only one equation which in the limiting cases reduces to the previously formulated models, namely
\begin{equation}
    \label{eq24}
    \dot{x}(t) = -V'(x) + \eta(t) + \varGamma(t) + \xi(t).
\end{equation}
Putting $D_D = 0$, one gets (\ref{eq1}). Similarly, in the absence of $\eta(t)$ ($D_P = 0$) eq. (\ref{eq12}) is restored. To obtain the average velocity $\langle v \rangle$ multiple realizations of the above system have to be simulated and then according to (\ref{eq18}) an ensemble average must be performed. The major disadvantage of this procedure is that a large number of sample paths must be generated to get statistically reliable results. The statistical error due to finite sampling is proportional to $1/\sqrt{N}$, where $N$ is the number of sample paths. However, there is also other type of error, called the systematic error, which emerges from the finite size of the time step $\Delta t$. The smaller the time step is, the larger number of sample trajectories must be simulated to get reliable results. In \cite{platen2010} it was pointed out that the following rule makes sense under suitable conditions: with a scheme of weak order of convergence $\beta$ it is sensible to increase $N$ at the order $\Delta t^{-2\beta}$. Apart from the above discussed shortcomings of computer simulation of stochastic differential equations another common challenge is to choose algorithms which are numerically stable for the task at hand. The precise definition of the numerical stability depends on the context but is typically related to the long term accuracy of an algorithm when applied to a given dynamics. Numerical errors such as round-off and truncation errors are often unavoidable during practical simulations. Therefore, it is important to choose a numerical scheme which does not propagate uncontrolled approximation errors. In particular, if quantities of interest concern the asymptotic long time limit features of the system such as the stationary average velocity $\langle v \rangle$ numerical stability is extremely important problem to obtain reliable results. In practice, the basic question on numerical stability should be almost always answered first when deciding which numerical scheme to use. Higher order of convergence is a secondary issue.
\begin{algorithm*}
    \caption{The weak order 2.0 predictor-corrector method to integrate $\dot{x}(t) = -V'(x) + \eta(t) + \varGamma(t) + \xi(t)$.}
    \label{predcorr}
    \begin{algorithmic}[1]
        \STATE $F_1 \leftarrow -V'(x_n)$
        \STATE $\bar{x}_{n+1} \leftarrow x_n + F_1\Delta t + \Delta\xi_n$
        \STATE $F_2 \leftarrow -V'(\bar{x}_{n+1})$
        \STATE Predictor: $\bar{x}_{n+1} \leftarrow x_n + \frac{1}{2}(F_1 + F_2)\Delta t + \Delta\xi_n$
        \STATE $F_2 \leftarrow -V'(\bar{x}_{n+1})$
        \STATE Corrector: $x_{n+1} \leftarrow x_n + \frac{1}{2}(F_1 + F_2)\Delta t + \Delta\xi_n + \Delta\eta_n + \Delta\varGamma_n$
    \end{algorithmic}
\end{algorithm*}

Taking all this into account, we decided to employ the weak order 2.0 predictor-corrector method \cite{platen2010} to simulate the stochastic dynamics given by (\ref{eq24}). According to the most comprehensive reference \cite{platen2010} it remains accurate, even when using relatively large time step sizes. Furthermore, this method is only slightly computationally more expensive than corresponding explicit schemes, as it uses the same random variables. Predictor-corrector algorithm is similar to implicit methods but does not require the solution of an algebraic equation at each step. Therefore it offers good numerical stability properties which it inherits from the implicit counterparts of its corrector. Moreover, the stated order of convergence is always defined for a worst case scenario, and higher order is often possible for specific classes of stochastic differential equations, e.g. those with additive noise as it is in (\ref{eq24}). In the above algorithm one first computes at each step the predicted approximate value $\bar{x}_{n+1}$ and afterwards the corrected value $x_{n+1}$. The difference $\chi = \bar{x}_{n+1} - x_{n+1}$ between these values provides information about the local approximation error. This quantity can be utilized to introduce a dynamic time step size control during the calculations to increase the efficiency of simulation.
\begin{algorithm*}
    \caption{The simplified scheme to generate the Gaussian increment $\Delta\xi_n$.}
    \label{gaussian}
    \begin{algorithmic}[1]
        \STATE $G = \sqrt{2D_G}$
        \STATE $\varphi \leftarrow U(0,1)$
        \IF {$\varphi \leq 1/6$}
            \STATE $\Delta\xi_n \leftarrow -G\sqrt{3\Delta t}$
        \ELSIF {$\varphi > 1/6 \wedge \varphi \leq 1/3$}
            \STATE $\Delta\xi_n \leftarrow G\sqrt{3\Delta t}$
        \ELSE
            \STATE $\Delta\xi_n \leftarrow 0$
        \ENDIF
    \end{algorithmic}
\end{algorithm*}

For weak convergence we only need to approximate the probability distribution induced by the process $x(t)$. Therefore, we can replace the Gaussian increments $\Delta \xi_n$ appearing in Algorithm \ref{predcorr} by other random variables with similar moment properties. We obtain a simpler scheme by choosing more easily generated random numbers, e.g. 
\begin{equation}
\label{eq25}
    \mbox{Pr}\{\Delta \xi_n = \pm G\sqrt{3\Delta t}\} = \frac{1}{6}, \quad \mbox{Pr}\{\Delta \xi_n = 0\} = \frac{2}{3},
\end{equation}
where $G = \sqrt{2D_G}$ is thermal noise intensity factor. It leads to the simplified weak order 2.0 predictor-corrector method \cite{platen2010}. This variant of the algorithm can be tens of times faster than the typical version based on the Gaussian random variables. The increase in efficiency reflects the fact that the simplified scheme requires only generation of uniformly distributed random number and therefore is computationally less expensive than its Gaussian counterpart. The loss in accuracy, which is due to the use of multi-point random variables, is for typical parameters below a few percent in terms of relative errors. However, in some particular cases the accuracy of simplified scheme can be superior to that built on Gaussian random numbers. Moreover, in general, one can say that the simplified algorithm often increases numerical stability \cite{platen2010}.
\begin{algorithm*}
    \caption{The algorithm to calculate the Poissonian increment $\Delta \eta_n$.}
    \label{poissonian}
    \begin{algorithmic}[1]
        \STATE $a \leftarrow \sqrt{\lambda D_P}$
        \IF {$s \leq 0$}
            \STATE $\zeta \leftarrow \sqrt{D_P/\lambda}$
            \STATE $\varphi \leftarrow U(0,1)$
            \STATE $s \leftarrow [-(1/\lambda)\ln{(1-\varphi)}/\Delta t]$
            \STATE $\varphi \leftarrow U(0,1)$
            \STATE $z \leftarrow -\zeta\ln{(1-\varphi)}$
            \STATE $\Delta\eta_n \leftarrow z - a\Delta t$
        \ELSE
            \STATE $s \leftarrow s - 1$
            \STATE $\Delta\eta_n \leftarrow -a\Delta t$
        \ENDIF
    \end{algorithmic}
\end{algorithm*}

To incorporate the white Poissonian noise (\ref{eq3}) into the presented scheme we first note that each sample realization of the process $x(t)$ jumps by a random step from the prescribed distribution only at the Poisson time $t_i$. During the times between random kicks $t_{i-1} < t < t_i$ the stochastic dynamics given by (\ref{eq24}) reduces to
\begin{equation}
    \label{eq26}
    \dot{x}(t) = -V'(x) - a  + \varGamma(t) + \xi(t).
\end{equation}
Then it is sufficient to integrate it by employing Algorithm \ref{predcorr} with $\Delta \eta_n = -a\Delta t$. The difference between subsequent Poissonian times $s_i = t_i - t_{i-1}$ is exponentially distributed
\begin{equation}
    \label{eq27}
        \psi_{\lambda}(s) = \lambda\theta(s)e^{-\lambda s},
\end{equation}
hence it can be obtained by the standard transformation $s=-\ln(u)/\lambda$
of the random number $u$ uniformly distributed on $(0,1)$. 
In our case the random amplitudes $\{z_i\}$ are also exponential random variates so they can be generated in the similar way. At each Poisson time $t_i$ the kick gives rise to a finite jump $z_i$ in $x(t)$. Then the increment is given as $\Delta \eta_n = z_i - a\Delta t$. An obvious shortcoming of this method is the fact that it requires variable time steps which must be adapted to each sequence of Poisson times $t_i$. It needs to be emphasized that presented jump-adapted simplified predictor-corrector approximation \cite{platen2010} becomes computationally demanding when the frequency $\lambda$ of the Poissonian kicks is high, as then the integration time step must be reduced. However, a jump-adapted discretization allows for use of various schemes for the pure drift and diffusion part of the dynamics between subsequent Poisson times. In particular, higher order jump-adapted weak schemes avoid multiple stochastic integrals that involve the white Poissonian noise. For this reason they are relatively simple and can be implemented without much effort. There exist numerical methods that use a constant time step \cite{kim2007, grigoriu2009, platen2010} but usually their weak order of convergence is lower. Higher order regular schemes are possible to derive, however they are really complicated even for the simplest stochastic differential equations \cite{platen2010}.
\begin{algorithm*}
    \caption{The algorithm to generate the dichotomous increment $\Delta \varGamma_n$.}
    \label{dichotomous}
    \begin{algorithmic}[1]
        \IF {$s \leq 0$}
            \IF {$d = 0$}
                \STATE $d \leftarrow 1$
                \STATE $\varphi \leftarrow U(0,1)$
                \STATE $s \leftarrow [-(1/\mu_b)\ln{(1-\varphi)}/\Delta t]$
                \STATE $\Delta \varGamma_n \leftarrow F_b\Delta t$
            \ELSE
                \STATE $d \leftarrow 0$
                \STATE $\varphi \leftarrow U(0,1)$
                \STATE $s \leftarrow [-(1/\mu_a)\ln{(1-\varphi)}/\Delta t]$
                \STATE $\Delta \varGamma_n \leftarrow -F_a\Delta t$
            \ENDIF
        \ELSE
            \STATE $s \leftarrow s - 1$
            \IF {$d = 0$}
                \STATE $\Delta \varGamma_n \leftarrow -F_a\Delta t$
            \ELSE
                \STATE $\Delta \varGamma_n \leftarrow F_b\Delta t$
            \ENDIF
        \ENDIF
    \end{algorithmic}
\end{algorithm*}

Basically, the same reasoning can be repeated to include the dichotomous noise in the employed numerical algorithm, see \cite{palleschi1988}. For definiteness we consider only a case of the $-F_a$ state. The sojourn times for staying in this state without flipping is governed by the exponential distribution
\begin{equation}
    \label{eq29}
        \psi_{\mu_a}(s) = \mu_a\theta(s)e^{-\mu_a s}.
\end{equation}
During this period the equation (\ref{eq24}) is reduced to
\begin{equation}
    \label{eq30}
        \dot{x}(t) = -V'(x) + \eta(t) - F_a + \xi(t)
\end{equation}
and can easily be integrated by using the previously outlined algorithms with $\Delta \varGamma_n = -F_a\Delta t$. For the next interval sampled from the above distribution with the transition rate $\mu_b$ the dichotomous noise stays in the $F_b$ state. Although presented procedure is simple due to the structure of the dichotomous noise it should be emphasized that this method causes serious problems when is employed to stiff systems with the noise of a short correlation time. It can be remedied by use of the regular scheme which has a constant time step \cite{kim2006}. However, such algorithm is based on the Euler method 
and therefore is of lower order of convergence.
\section{Best practices guide}
\label{sec4}
This manual is a guide how to properly accelerate numerical solution of stochastic differential equations by harvesting the power dormant in modern NVIDIA GPU. It presents established parallelization and optimization techniques which help to obtain maximum performance from CUDA. Performance optimization revolves around three basic strategies \cite{cuda}:
\begin{enumerate}
    \item Optimize memory usage to achieve maximum memory throughput;
    \item Maximize parallel execution to achieve maximum utilization;
    \item Optimize instruction usage to achieve maximum instruction throughput.
\end{enumerate}
Based on the recommendations contained in \cite{cuda} we now will detail on each of these levels of optimization and comment how it is realized in the code supplemented to this paper.
\begin{algorithm*}
    \caption{A CUDA kernel to simulate the models given by (\ref{eq1}) and (\ref{eq12}).}
    \label{kernel}
    \begin{algorithmic}[1]
        \STATE local $i \leftarrow blockIdx.x \cdot blockDim.x + threadId.x$
        \STATE load $x_i$, system parameters, the noise and RNG state from global memory and store them locally in the set of registers
        \FOR {$j = 1$ \TO samples}
            \STATE advance $x_i$ by $\Delta t$ using Algorithm \ref{predcorr}
        \ENDFOR
        \STATE save $x_i$, the noise and RNG state to global memory
    \end{algorithmic}
\end{algorithm*}

\subsection{Memory optimization}
Memory optimizations are the most important area for performance. Its hierarchy described in Sec. \ref{sec2} instantly suggest a strategy to follow which can be summarized as: move as much data as possible to the fastest kind of memory and keep it there as long as possible.
\begin{itemize}
    \item \emph{Minimize data transfer between the host and the device.} \\ Theoretically, a single CUDA kernel can be responsible for the entire numerical simulation. This results in just two data transfers, one before and after the simulation. The intermediate data structures are created in device memory, operated by the device and destroyed without ever being touch by the host. However, GPUs installed in desktop computers often have run time limit on kernels. In order to avoid the possible exceeding of this barrier it is reasonable to split main kernel into smaller parts, each calculating \emph{samples} number of time steps in a single invocation.
    \item \emph{Make global memory accesses coalesced whenever possible.} \\ Beside organization of threads into blocks, there is also further arrangement into warps. These are the actual number of threads that gets calculated in SIMT. Global memory accesses by threads of a warp are coalesced by the device into as few as one transaction when certain requirements are met. In order to attain good bandwidth utilization we employed a natural parallelism - each realization of the stochastic process is calculated in separate thread. Moreover, we run the simulation kernel in grid of one-dimensional blocks with typically 256 threads for each one of them. The latest commodity of NVIDIA GPUs have warp size set to 32 so such a fine-grained parallelism ensures that all global memory accesses are coalesced.
    \item \emph{Minimize data transfer between global memory and the device.} \\ This goal is achieved by maximizing utilization of on-chip memory. In particular, it is recommended to use the fastest type of memory, i.e. the set of registers. At the beginning of each kernel invocation essential variables are loaded from global memory to registers and stay there for all calculations. The reverse operation is performed at the end of kernel execution.
\end{itemize}
\subsection{Execution configuration optimizations}
To maximize utilization the program should be structured in such a way that it exposes as much parallelism as possible and keep the device busy most of the time.
\begin{itemize}
    \item \emph{Maintain sufficient occupancy of the device.} \\ Occupancy is the ratio of the number of active warps per multiprocessor to the maximum number of possible active warps. It should be maintained at the appropriate level because thread instructions are executed sequentially in CUDA and therefore executing other warps when one warp is busy is the only way to hide latency. In order to increase the number of active threads in a straightforward manner we could divide the planned computational work between more threads. However, the application of this idea would unnecessary complicate the code. Therefore we decided not to do it. The reason for this lies in the peculiarities of the models (\ref{eq1}) and (\ref{eq12}) where one is usually interested in calculating certain quantities for the whole range of the system parameters. It is much more reasonable to structure the code in the clever way, such that the stochastic dynamics is solved for multiple values of the system parameters in a single kernel call. This procedure allows for very efficient exploration of the parameter space.
    \item \emph{Thread and block heuristics.} \\ The number of threads per block should be a multiple of 32, as this provides the optimal computing efficiency and facilitates coalescing.
\end{itemize}
\subsection{Instruction optimization}
To maximize instruction throughput the program should mainly make as large as possible the usage of simple arithmetic instructions which often have their own hardware accelerated implementation on CUDA GPU.
\begin{itemize}
    \item \emph{Use single-precision arithmetic whenever it is possible.} \\ On currently available GPUs the single precision operations are about an order of magnitude faster than the double precision ones. This fact is often considered as significant limitation for numerical calculation. However, the performance gap is becoming smaller for the newer devices. Anyhow, we have found that for the presented models it is not a problem and the use of single precision arithmetic does not significantly impact obtained results. To increase the precision without sacrificing performance, we introduced the folding of $x$ variable. As only $x\,\mbox{mod}\,L$ is relevant for the system dynamics, the device is instructed to operate only on $x\,\mbox{mod}\,L$, simultaneously keeping track of the rest of $x$ in an another variable. This kind of folding creates the possibility of avoiding the numerically pathological situations in which a small quantity is added to a much larger one. In single precision, such an operation is problematic if one of the quantities is more than 7 orders of magnitude larger than the other one. 
    \item \emph{Use the fast math library whenever speed trumps precision.} \\ CUDA GPUs provide an alternative hardware implementation of various transcendental function such as trigonometric functions, exponent, logarithm and so on. In particular, they are hardware-accelerated by the Special Functions Unit of the scalar microprocessor when their argument do not fall within a narrow range specified for each of them. This leads to an increase of performance at the cost of precision. However, we have found that use of these functions does not degrade the results of the simulation if the above precautions are taken.
\end{itemize}
Of course, apart from the above listed optimizations there is still some room for a further performance improvement. In particular, it includes utilization of the parallel algorithms, e.g. a sum reduction needed in the ensemble averaging. However, from our experience it follows that such an optimization matters significantly only for a case of a very large number of particles. Another possible improvement is connected with an avoidance of an instruction branching and a divergence within the same warp. Any flow control instruction (if, switch, do, for, while) can significantly affect the instruction throughput by causing threads of the same warp to diverge; that is, to follow different execution paths. 
It is generally not an easy task to avoid such problems in designing code which suppose to be as generic as possible, especially for a programming language like CUDA C.
Nevertheless this can be remedied by use of template-based Run-Time Code Generation (RTCG) \cite{januszewski2014} which is utilized to automatically generate optimized code for a specific system of stochastic differential equations. Moreover, it can also provide some isolation from low-level hardware details of rapidly evolving GPUs, greatly saving time and increasing the productivity of a programmer. Finally, while the computational power available in a single GPU is impressive, one may naturally run a distributed simulation using multiple GPUs which can be physically located in a single computer or in multiple machines connected over a network \cite{januszewski2014}.

\section{Validation}
\label{sec6}
In order to validate our implementation of the presented algorithms, we performed simulations for the two previously discussed models and compared our results with the known analytical counterparts.

\begin{figure}[h]
    \centering
    \includegraphics[width=0.45\linewidth]{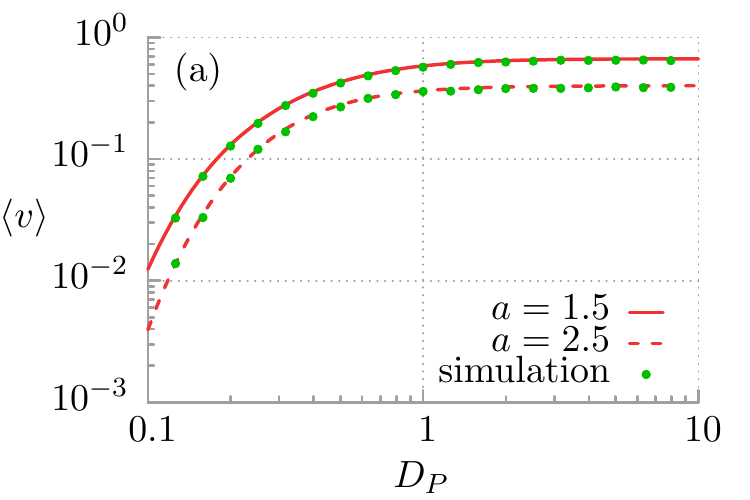}
    \includegraphics[width=0.45\linewidth]{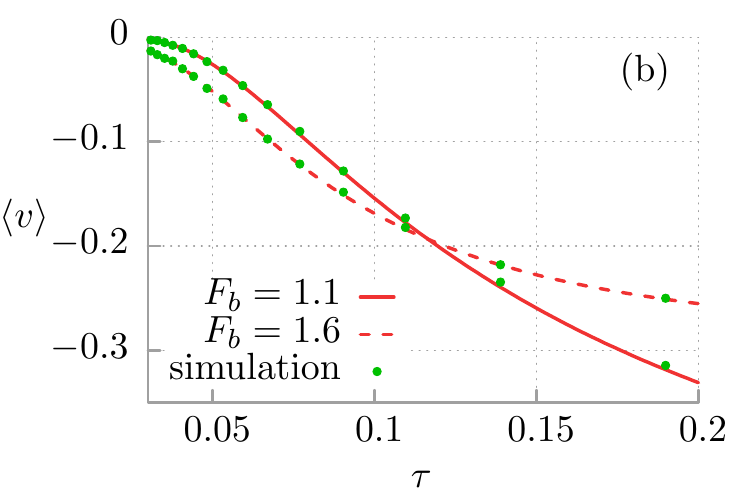}
    \caption{Comparison between the simulated (points) and theoretical (curves) asymptotic long time stationary average velocity $\langle v \rangle$ of the Brownian particle described by the models (\ref{eq1}) and (\ref{eq12}). Panel (a): the average velocity $\langle v \rangle$ versus the white Poissonian noise intensity $D_P$ for the two values of the asymmetry parameter $a = \sqrt{\lambda D_P} = 1.5$ (solid line) and $a = 2.5$ (dashed line). 
    Panel (b): the dependence of the same transport characteristic as on (a) on the dichotomous noise correlation time $\tau$ is depicted for $F_b = 1.1$ (solid line), $F_b = 1.6$ (dashed line) and fixed $F_a = 3$. In both panels the Gaussian noise intensity $D_G$ is set to zero.} 
    \label{fig3}
\end{figure}

The tests were run in a single precision using the CUDA Toolkit 5.5 on a 64-bit Linux system. Enabling the full floating-point optimizations does not affect the results and therefore we set the \texttt{-O3 --use\_fast\_math} options of the NVIDIA compiler. In the case of the first model (\ref{eq1}) we carried out the simulations of $1024$ Brownian particles, each lasting of $10^3 \cdot (1/\lambda)$ dimensionless units of time. Here, we note that $1/\lambda$ is the natural characteristic time scale for the studied system. It is reasonable not to specify the time step directly. Instead, it should be dynamically adapted to the chosen characteristic time scale.
We provided the $spp = 400$ parameter which stands for the \emph{steps per period}. This number together with the spiking rate of the white Poissonian noise $\lambda$ was used to calculate the suitable step size $\Delta t = (1/\lambda)/400$. Each kernel invocation advanced the simulation by $samples = 4000$ time steps. We estimated the asymptotic long time stationary average velocity $\langle v \rangle$ from the results of numerical calculations in the following way
\begin{equation}
    \label{eq31}
        \langle v \rangle \approx \langle v \rangle_{sim} = \left \langle \frac{x(t_f) - x(t_i)}{t_f - t_i} \right \rangle,
\end{equation}
where $\langle \cdot \rangle$ stands for the ensemble average. The time average is performed over a large, but finite time span $t_f - t_i$. In order to discard the influence of the initial transient effects before the $\langle v \rangle$ is estimated we turned $10\%$ of the $t_f$ down, i.e. set $t_i=t_f/10$. The simulations show the perfect agreement with the analytical predictions published by {\L}uczka et al. \cite{luczka1995} (see Fig. \ref{fig3}a).

For the second model (\ref{eq12}), we simulated the same number of Brownian particles. However, for the dichotomous dynamics there are two natural characteristic time scales, which correspond to the mean waiting time in the first $\tau_a = 1/\mu_a$ and the second $\tau_b = 1/\mu_b$ state of the two--state noise. In all calculations we took as a reference the smaller one. Each trajectory evolved until $10^3$ dimensionless characteristic time scales of the system. The \emph{spp} parameter was fixed to $200$ and each kernel call calculated $samples = 2000$ time steps. Similarly to the previous case initial $10\%$ of the simulation time was dropped down for the estimation of $\langle v \rangle$ due to the possible transients effects. The comparison between the simulated and theoretical results presented in \cite{kula1996} is illustrated in the Fig. \ref{fig3}b. Again, the perfect agreement is found.
\section{Performance evaluation}
\label{sec7}
In this section the performance evaluation of the discussed numerical solution of the models (\ref{eq1}) and (\ref{eq12}) is presented and compared to the corresponding CPU C implementation. The latter was obtained from the original CUDA source code with the minimum changes possible and the proper replacement of all of the parallel parts.

GCC 4.8.2 was used to compile the CPU version. Here, we present benchmarks only for the two limiting cases: 
highly optimized single precision version (\texttt{nvcc -O3 --use\_fast\_math}, \texttt{gcc -O3 -ffast-math}) and the double precision counterpart with the default compiler flags. All tests were conducted on the same 64-bit Linux machine using the following hardware:
\begin{itemize}
    \item CPU: Intel Core i7-3930K @ 3.2GHz and 32 GB RAM. A single core was used for the simulation.
    \item GPU: NVIDIA GeForce GTX-TITAN BE installed in a system with the above described CPU.
\end{itemize}
\begin{figure}[t]
    \centering
    \includegraphics[width=0.45\linewidth]{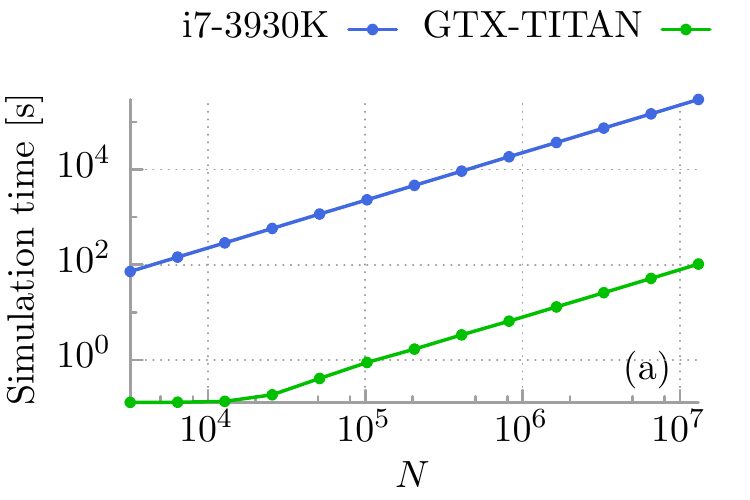}
    \includegraphics[width=0.45\linewidth]{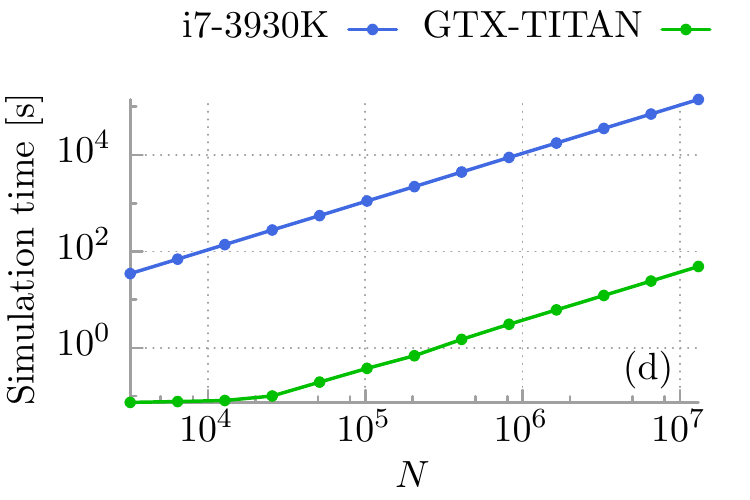}\\
    \includegraphics[width=0.45\linewidth]{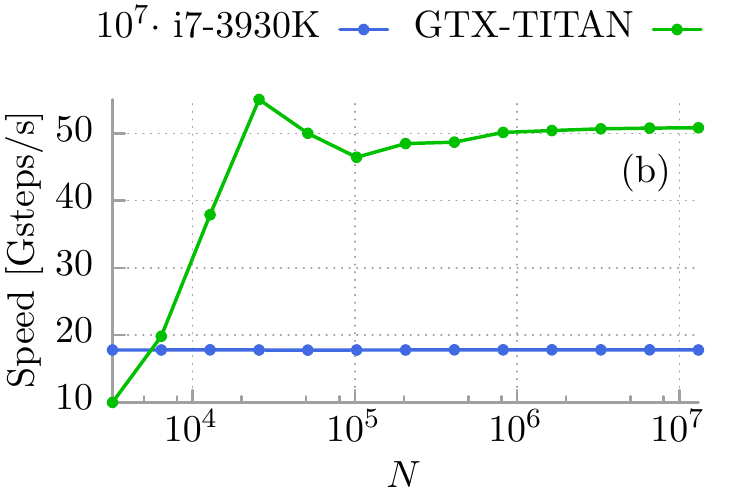}
    \includegraphics[width=0.45\linewidth]{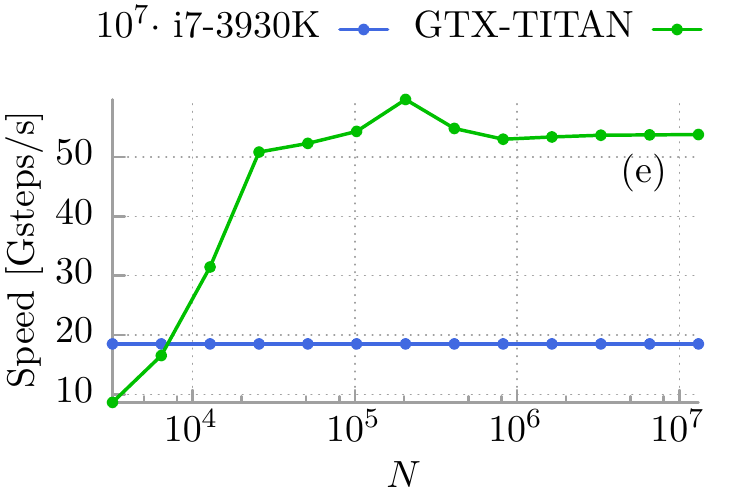}\\
    \includegraphics[width=0.45\linewidth]{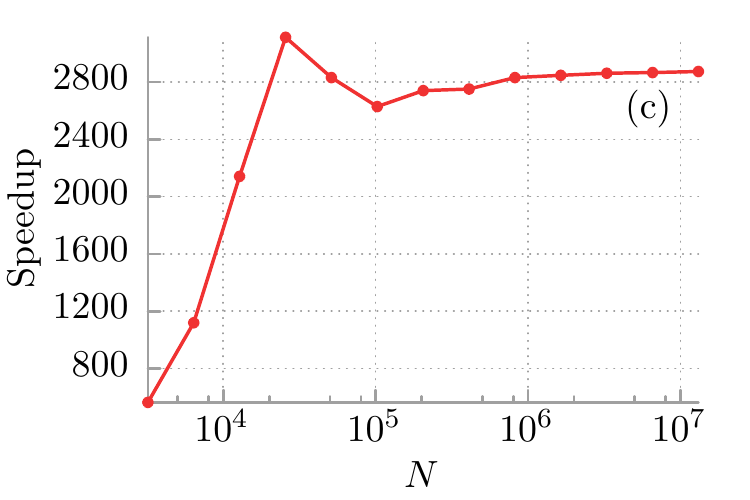}
    \includegraphics[width=0.45\linewidth]{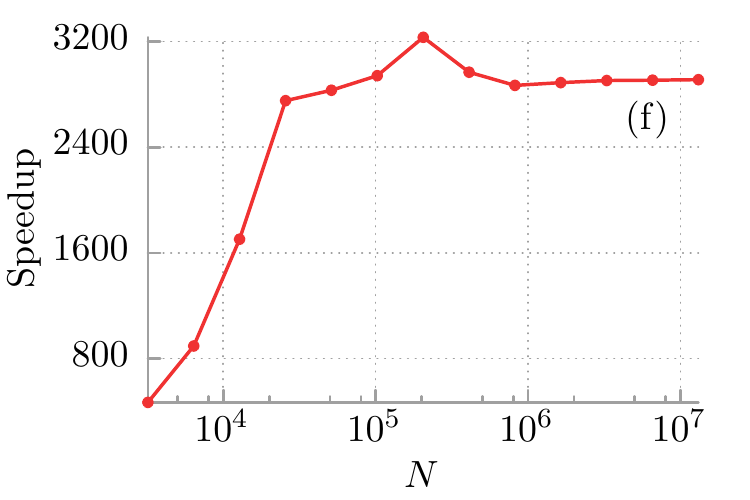}
    \caption{Single precision performance estimates for the programs \emph{poisson} and \emph{dich} as a function of the number of particles $N$. Left and right panels correspond to the first and the second model, respectively. The top plots present the total simulation time in seconds. The middle ones illustrate the speed estimation (\ref{eq32}). 
    The bottom panels depict the GPU speedup (\ref{eq33}).}
    \label{fig4}
\end{figure}
\begin{figure}[t]
    \centering
    \includegraphics[width=0.45\linewidth]{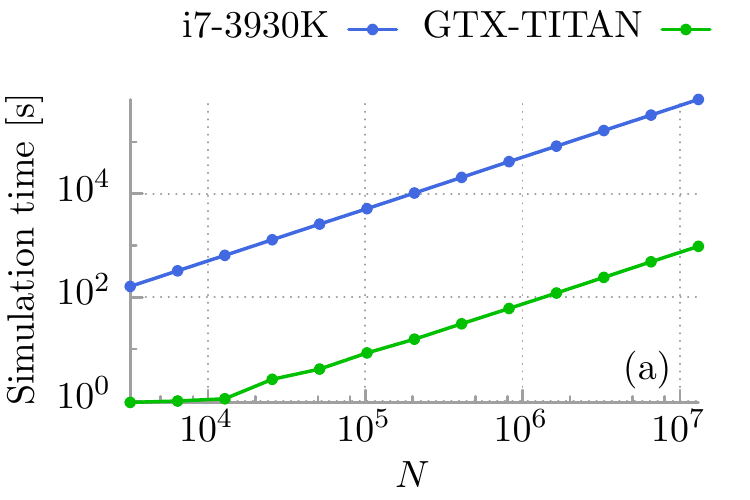}
    \includegraphics[width=0.45\linewidth]{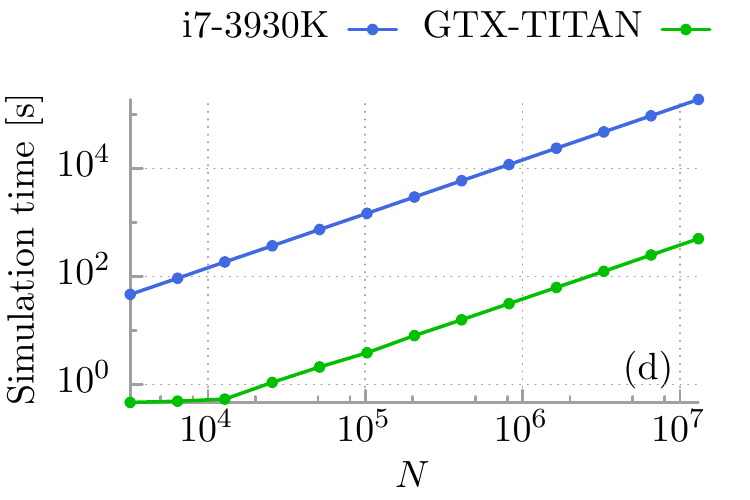}\\
    \includegraphics[width=0.45\linewidth]{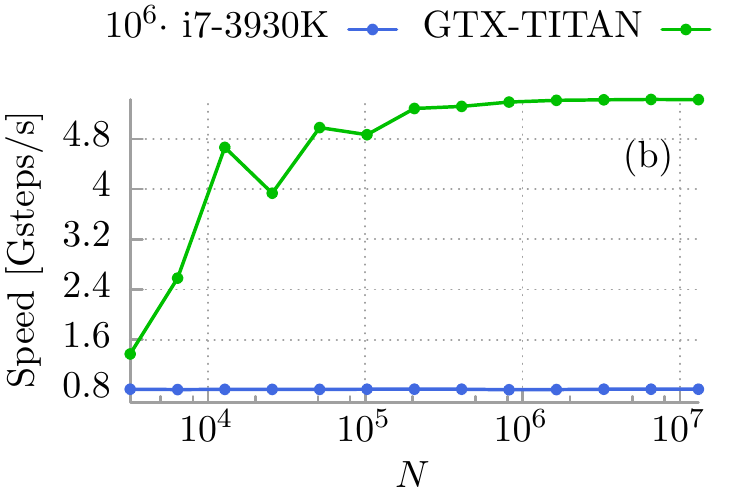}
    \includegraphics[width=0.45\linewidth]{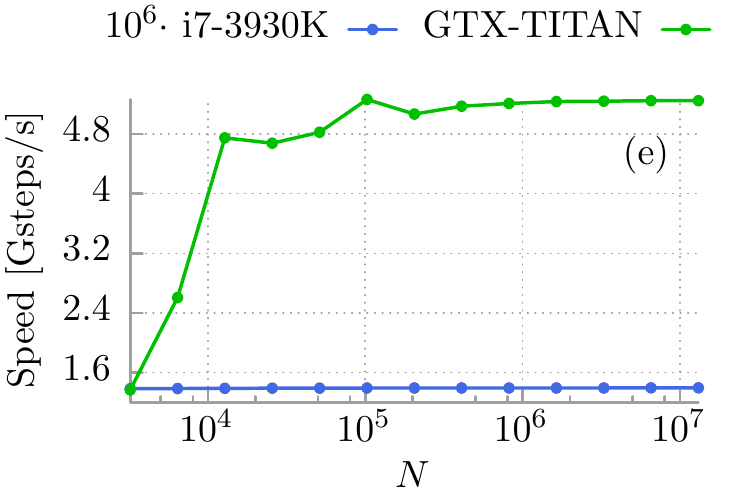}\\
    \includegraphics[width=0.45\linewidth]{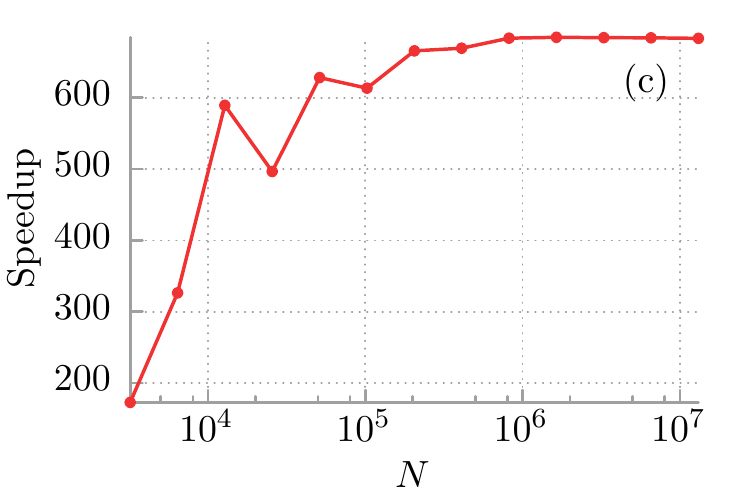}
    \includegraphics[width=0.45\linewidth]{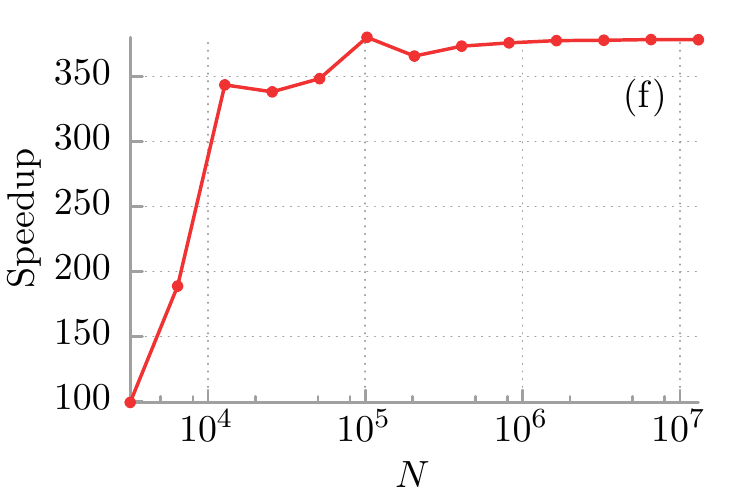}
    \caption{Double precision performance estimates for the programs \emph{poisson} and \emph{dich} as a function of the number of particles $N$. For the rest of the details see Fig. \ref{fig4}.}
    \label{fig5}
\end{figure}
Fig. \ref{fig4} and Fig. \ref{fig5} illustrate the performance of the discussed numerical solutions as a function of the total number of the Brownian particles $N$ for the single and double precision variant, respectively. The performance is defined by means of three measures: a simulation time given in seconds, a speed defined as
\begin{equation}
    \label{eq32}
    \mbox{Speed} = 10^9 \cdot \frac{\mbox{total number of time steps}}{\mbox{simulation time}}
\end{equation}
and finally an actual speedup expressed in the following way
\begin{equation}
    \label{eq33}
    \mbox{Speedup} = \frac{\mbox{CPU simulation time}}{\mbox{GPU simulation time}}.
\end{equation}

Our tests show that in the case of the highly optimized single precision arithmetic the speedups of the astonishing order of about 3000 are possible for the models (\ref{eq1}) and (\ref{eq12}). Even when the entire simulation is carried by the double precision variables it is still feasible to achieve an amazing performance gain of the order of about 700 and 400 as compared to the CPU implementation. However, this performance profit is clearly dependent on the number of Brownian particles used in the simulation. Panels (c) and (f) of Fig. \ref{fig4} and Fig. \ref{fig5} depict that initially it increases with the number of simulated trajectories and then saturates around $N = 10^5$. This point corresponds to the situation when the computational resources of the GPU are fully utilized. Moreover, for each program there is an ideal number of Brownian particles when the measured speedup is largest. It is because for these points the device occupancy is reasonable high and at the same time the so called register pressure is optimal. Register pressure occurs when there are not enough registers available for a given task. Although each MP contains thousands of 32-bit registers these are partitioned among concurrent threads and therefore one should keep the pressure at the appropriate level \cite{NugMes2010}.

Overall, by comprehensive profiling of our codes with NVIDIA Visual Profiler we found that the limiting factor of the simulation kernel execution is the GPU instruction throughput. However, there is not much room for improvement in this field because wherever it is possible we used alternative hardware version of transcendental functions. Therefore we conclude that our implementation of the task at hand is very close to optimum. Since the GPU that was used to benchmark our code is neither the most advanced nor the fastest NVIDIA GPU available on the market when this paper is written, one can basically expect even better speedups. Moreover, there is still possibility to run a distributed simulation using multiple GPUs which for convenience can also be located in a single machine. In such situation, achieving the performance gain of unbelievable order $10^4$ would be just a piece of cake.
\section{Conclusions}
\label{sec8}
In this paper we have presented an updated and extended step by step guide on how to numerically solve the stochastic differential equations with both Gaussian and non-Gaussian fluctuations, especially in the form of the white Poissonian noise and the dichotomous process. In particular, we have properly harvested the power dormant in the latest commodity of Graphics Processing Units and demonstrated the possibility of achieving speedups of the order as large as about 3000 when compared to the same implementation executed by the single CPU core. This is thanks to the inherent parallel nature of the Monte Carlo integration of the stochastic differential equations. However, in order to achieve such astonishing performance gain one has to carefully redesign the known algorithms. Hopefully, with the help of our work this task will become much easier.

Finally, it is certain that the potential of the General Purpose computing on Graphics Processing Units still awaits to be fully revealed. Nevertheless, even now it is safe to say that it opens a completely new chapter in the history of high performance computing.
\section*{Acknowledgments}
This work was supported in part by the MNiSW program ”Diamond Grant” (J.S.).

\end{document}